# In-situ Magnesium Diboride Superconducting Thin Films grown by Pulsed Laser Deposition

G.Grassano, W.Ramadan, V.Ferrando, E.Bellingeri, D.Marré, C.Ferdeghini, G.Grasso, M.Putti, A.S.Siri
INFM, Dipartimento di Fisica, Via Dodecaneso 33, 16146 Genova, Italia

P.Manfrinetti, A Palenzona
INFM, Dipartimento di Chimica e Chimica Industriale, Via Dodecaneso 31,16146 Genova, Italia

A.Chincarini
INFN Dipartimento di Fisica, Via Dodecaneso 33, 16146 Genova, Italia

## Abstract

Superconducting thin films of $MgB_2$ were deposited by Pulsed Laser Deposition on magnesium oxide and sapphire substrates. Samples grown at 450ºC in an argon buffer pressure of about $10^{-2}$ mbar by using a magnesium enriched target resulted to be superconducting with a transition temperature of about 25 K. Film deposited from a $MgB_2$ sintered pellet target in ultra high vacuum conditions showed poor metallic or weak semiconducting behavior and they became superconducting only after an ex-situ annealing in Mg vapor atmosphere. Up to now, no difference in the superconducting properties of the films obtained by these two procedures has been evidenced.

The recent outstanding discovery of 40K superconductivity in $MgB_2$ [1, 2] is a breakthrough that stimulated a worldwide excitement in the scientific community renewing the interest on superconductivity. This material exhibits a lot of intriguing properties: it is the compound with the highest $T_c$ among non-oxides materials and first reports indicate a phonon mediated mechanism for the superconductivity making $MgB_2$ also the BCS materials with the highest $T_C$ [3]. Furthermore, the grains boundaries have not dramatic effects on the critical current densities, being of its coherence length longer than those of HTSC. The reported properties, the very simple structure and the commercial availability of this material make $MgB_2$ a favorite candidate for large scale and electronic applications and suggest the possibility of further progress in technology based on superconducting materials. The main limitation for the application seems to be related to the considerable small value of the irreversibility field (about 7 Tesla at liquid helium temperature) [4].

High quality epitaxial thin film are needed to implement a new class of electronic devices. Actually, thin film deposition of this material is a quite difficult task due to the high volatility of Mg, so that $MgB_2$ growth seems impossible in vacuum conditions. This necessity imposes severe restrictions on deposition systems and, up to now, superconducting thin films have been produced only via ex-situ annealing in magnesium atmosphere [5, 6]. Very recently, an in situ annealing has been proposed [7] to react, after the deposition, stoichiometric or Mg rich mixture of magnesium and boron, but, up to now, no thin films have shown superconductivity as grown. For electronic applications, however, the annealing procedure is not a suitable technique to produce superconducting thin film and an in situ deposition of $MgB_2$ is necessary in order to allow the integration of layers of different materials as in devices or junctions.

In this paper, we report on the deposition of superconducting $MgB_2$ films by Pulsed Laser Deposition (PLD). By using a stoichiometric superconducting target, we performed a systematic study on the influence of the growth temperature on the electrical properties of thin film. Furthermore, for the first time, by using a magnesium rich target and by depositing in a argon buffer gas pressure, we were able to deposit superconducting samples with transition temperature of about 25 K. This latter method, even if still not optimized, open new perspectives for realizing $MgB_2$ based devices.

## Ex situ deposition

For the ex situ procedure we used $MgB_2$ sintered target prepared by direct synthesis from the elements: Mg ( 99.999 wt.% purity), in form of fine turnings, and crystalline B (99.7 wt.% purity), 325 mesh, were well mixed together and closed by arc welding under pure argon into outgassed Ta crucibles which were then closed in quartz ampoules under vacuum. The samples were slowly heated up to 950 °C and maintained at this temperature for 1 day.

The target was then obtained by sintering these powders in pellets at 1100 °C for 3 days again in Ta containers, sealed under argon, and in silica tubes closed under vacuum.

A transition temperature $T_c$ of 39K was determined from magnetization and resistivity measurements, and a residual resistivity ratio RRR=R(300K)/R(40K) of 20 was determined using a four probe resistive method.

$MgB_2$ thin film have been deposited by using a PLD deposition system described elsewhere [8].

The choice of the substrate is fundamental to drive the epitaxial growth of the film. The hexagonal $MgB_2$ symmetry suggests the use of $Al_2O_3$ (1102) substrates, that, despite a not negligible mismatch of the lattice parameters, present the same surface symmetry.

Furthermore, because a small amount of oxygen released by the substrate could hinder the formation of the correct phase, we used also MgO substrates which is very stable at high temperature. Moreover, due to its ionic character, MgO enables to obtain textured and epitaxial growth of other intermetallic superconductors like quaternary borocarbide compounds, even if the lattice mismatch is relevant [9].

Our first attempts to deposit $MgB_2$ films in high vacuum condition ($10^{-8} – 10^{-9}$ mbar) were carried out by scanning the different deposition parameters: mainly substrate temperature (ranging from room temperature up to 750°C) as well as the repetition rate (from 1 Hz to 30 Hz). Those attempts resulted in the formation of an amorphous phase from room temperature up to 300°C without any evidence of superconducting transition. By increasing the deposition temperature, the resistance changes from a poor metallic behavior at room temperature to a more marked semiconductor behavior.

At deposition temperatures higher than 350K, the films result nearly transparent. In these conditions, as evaluated by X-rays reflectivity measurements, a very small amount of material was deposited on the substrate. These feature is quite strange because even if the evaporation of Mg is expected, the boron must anyway remain on the substrate.

XPS and mass spectroscopy analyses of the residual gas (RGA) helped us to clarify this point: XPS measurement performed on films deposited at room temperature revealed the correct ratio between Mg and B, and also a remarkable presence of oxygen, which results chemically bounded with magnesium, but also with boron ( see the $B_2O_3$ peak near the $MgB_2$ peak in figure 1). To understand the effect of the temperature on the deposited phase, we heated in UHV up to 800°C one sample deposited at room temperature, performing both XPS, and RGA at temperature step of 1°C/minute . The relative intensity of the oxide-boride peaks changed with temperature in a complicate way, up to now not well understood, in figure 2 the partial pressure versus temperature for boron oxide is shown: at temperature higher than 360°C a relevant quantity of $B_2O_3$ is detected; this temperature is in agreement with the temperature at which the ablated species start evaporating from the substrate thus leading to the deposition of very small amount of material.

Regarding the origin of the oxides present in the thin film composition, XPS measurements performed on a piece of target revealed more or less the same quantity of oxide; this is put in evidence also by the green color of the plasma plume during the deposition. As suggested in [5], this color is related to optical emission by MgO, while the color of metallic magnesium is blue. The presence of oxygen in the target could be an obstacle for obtaining the formation of the superconducting phase of $MgB_2$ .

To crystallize the superconducting phase, an ex-situ annealing in magnesium vapor was carried out in an evacuated quartz tube at 650°C for 30 minutes followed by rapid quenching to room temperature. In figure 3 we report the film electrical resistivity before and after the thermal treatment in Mg vapor as a function of temperature for sample deposited at different temperature. We found out a clear relation between the resistance behavior before the annealing and the quality of the superconducting transition after the annealing procedure. In fact, as shown in figure 3a, when a poor metallic sample is annealed, we can observe superconductivity with an onset temperature of about 28.6K and a transition width (10%-90% variation of the resistance) of about 5.2K. In the case of an as grown slightly semiconducting sample (figure 3b), the same annealing procedure produced a film with worse superconducting properties ($T_c$=9.4K, $\Delta T_c$=2.6K), while we were not able to stabilize superconducting phase in the case of deeply semiconducting film deposited at high substrate temperatures. This could be due to the presence of very stable boron reach phase, as revealed by x-rays diffractometry, in agreement with the phase diagram of the system proposed in [3] and [10].

All the superconducting samples have a RRR slightly greater than 1, evidencing poor structural properties. Also the samples produced with similar ex situ annealing techniques, and presented up to now in literature, have a low RRR value. This could be an intrinsic limit of this method for realizing high quality film, because of an amorphous starting layer that could make difficult a good crystallization of the right phase.

For this reason, we have chosen also to explore a completely in situ growth of the correct phase, avoiding any kind of in annealing procedure, that could be promising for the deposition of high quality epitaxial thin films.

**In situ deposition**

In the in situ film deposition procedure, we tried to deal with the two problems forehead discussed i.e. magnesium evaporation and oxidation. A magnesium rich target, MgB, was prepared by mixing stoichiometric ratio of Mg powder (Alfa Aesar, purity 99.98%) and B powder (purity 99.9%) and pressed in the form of a pellet. In this way, the excess of magnesium, with respect to stechiometry 1:2, supplies an extra-source of this element, to simulate the effect of a magnesium atmosphere during the non-equilibrium deposition conditions. We would emphasize that the target was not thermally reacted or sintered but just a pellet of fining mixed and pressed powders. Other more Mg-rich compositions have been tried up to now without success.

The samples deposited in high vacuum condition and at substrate temperatures up to 600 °C resulted not superconducting.

To obtain in situ superconducting samples, it seems crucial the use of an argon buffer gas, in such a way to change the dynamics inside the plume during the plasma expansion. Increasing the background pressure, the color of the plume changes from green (the color of MgO, at pressure lower than $10^{-2}$mbar) to blue (the color of metallic Mg, at argon pressure in the range of $10^{-2}$mbar), and again to green (pressure higher than 6-8 $10^{-2}$mbar). This is a clear evidence that the chemical reactions and the oxidation processes in the plume are strongly affected by the presence of the buffer gas.

We have tested different argon background pressure (1mbar, $10^{-1}$mbar, and 2 x $10^{-2}$ mbar), in order to minimize the oxidation during the deposition. As suggested by the color of the plume, superconductivity was detected only in samples prepared at pressure of 2 $10^{-2}$mbar. The chosen deposition temperature was in this case 450°C and laser repetition rate was 30 Hz. The high repetition rate used provides an additional way to compensate the losses in Mg along with the magnesium rich target. The choice of the substrate temperature resulted in agreement with Zi-Kui Liu and al. [10] suggestions.

In figure 5 the resistance vs. temperature of the resulting superconducting film is reported and an onset transition temperature of 25K was recorded, the transition width was 2.1K.

In conclusion, we report on two different methods to obtain superconducting magnesium diboride thin films by Pulsed Laser Deposition. The first one consists in a room temperature deposition followed by an ex-situ annealing in a Mg vapor, the second one is, for the first time, a truly in situ growth, without annealing procedure. By this latter method, by using a magnesium rich target and by depositing in a argon buffer gas pressure, we were able to deposit in situ superconducting samples with transition temperature of about 25 K. A critical point of this procedure seems to be related with the argon partial pressure that determine the plasma dynamics. This process, although not yet optimized, could open new perspectives for realizing $MgB_2$ based devices.

Figure caption

Figure 1. XPS measurement performed on films deposited at room temperature: the $B_2O_3$ and $MgB_2$ peaks clearly reveal that oxygen is chemically bound also with boron.

Figure 2. Partial pressure versus temperature for boron oxide, as resulted from RGA: at temperature higher than 360°C a relevant quantity of $B_2O_3$ is detected.

Figure 3. (a) Resistance vs. temperature of a pre-anneling poor metallic sample, before and after the annealing at 650°C in a Mg-atmosphere and (b) resistance vs. temperature of a pre-anneling slightly semiconducting sample, before and after the annealing at 650°C in a Mg-atmosphere.

Figure 4. Resistance vs. temperature of a $MgB_2$ grown in situ starting from a Mg-rich, not reacted MgB target.

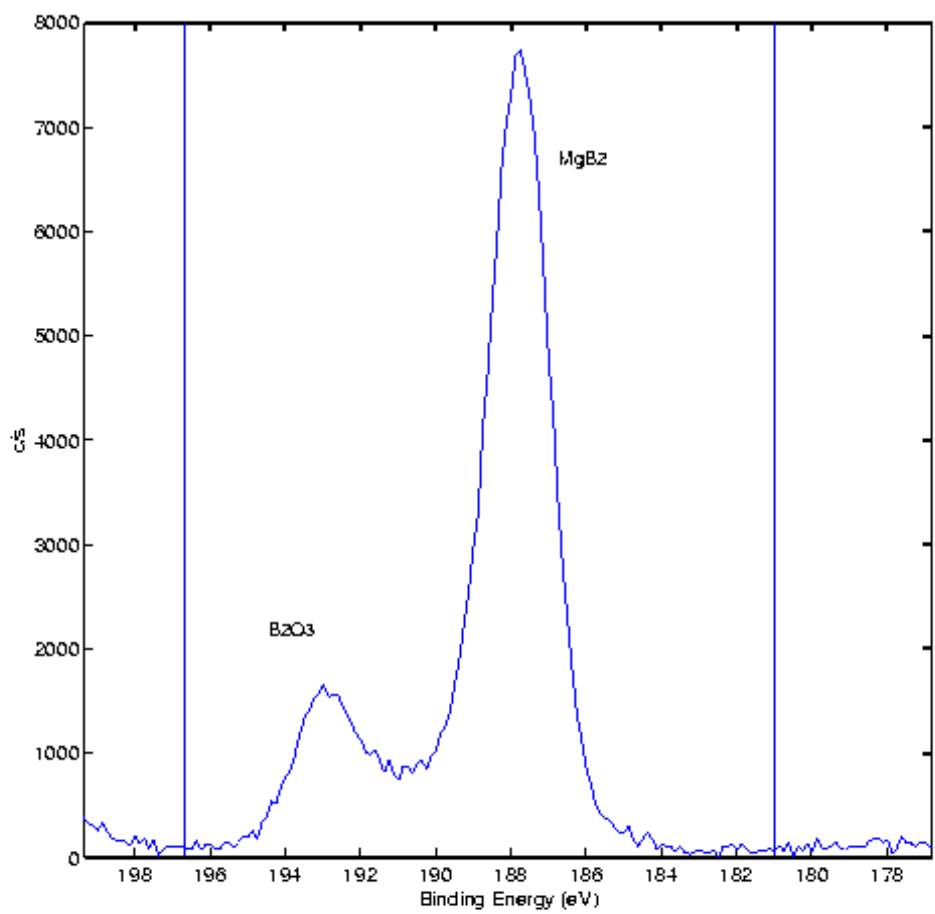

Figure 1

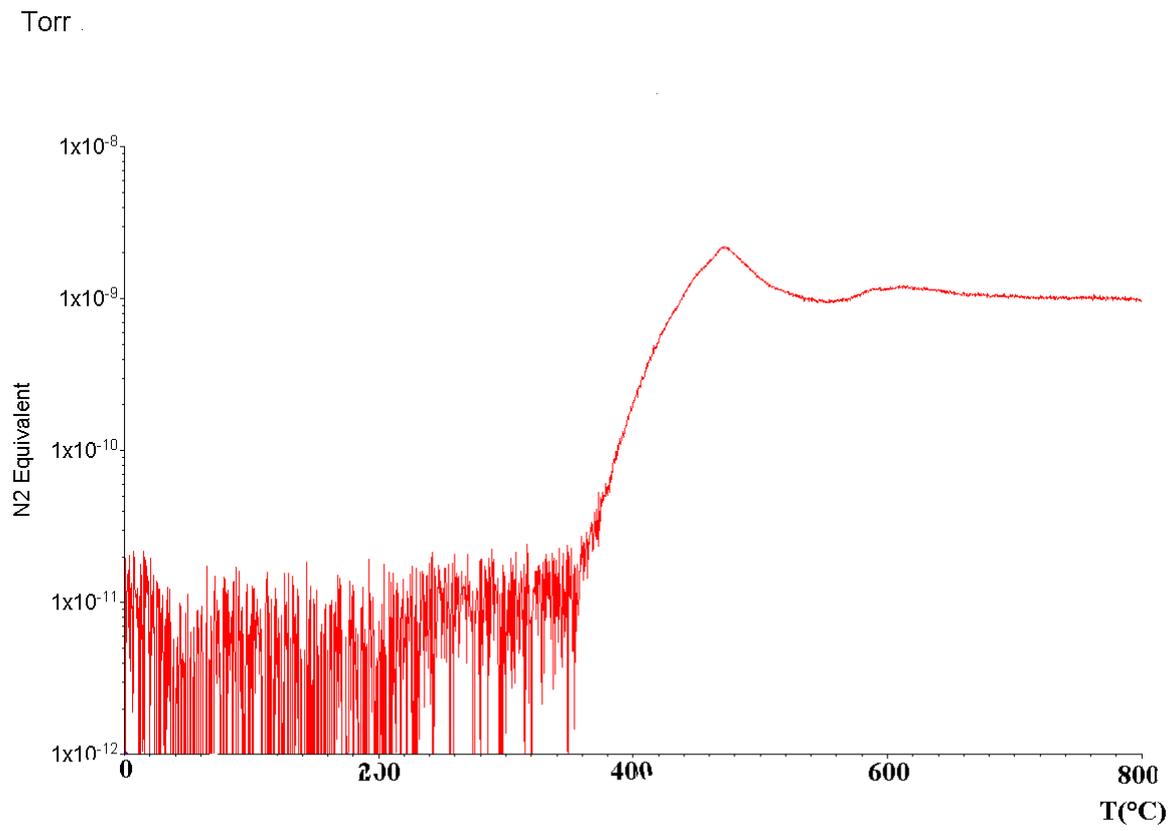

Figure 2

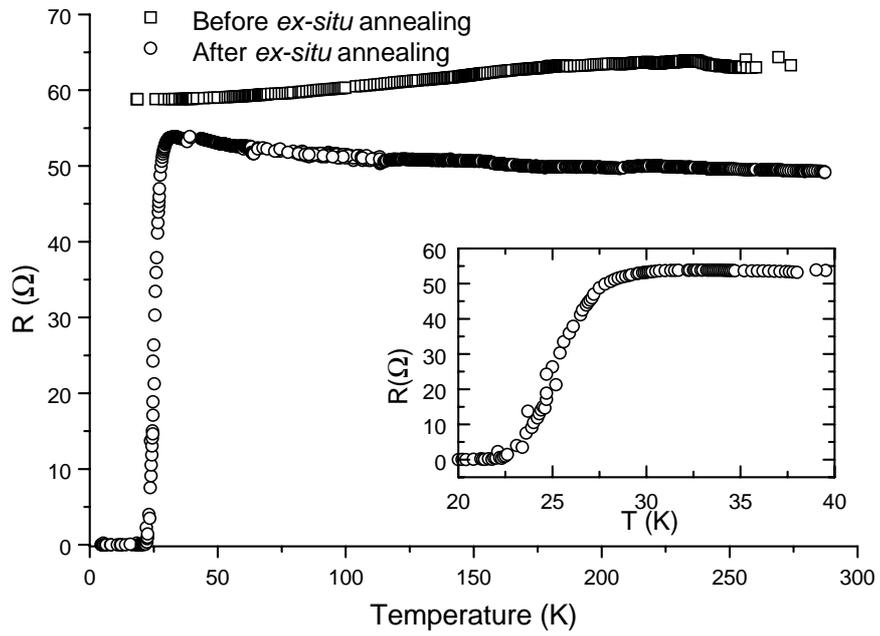

(a)

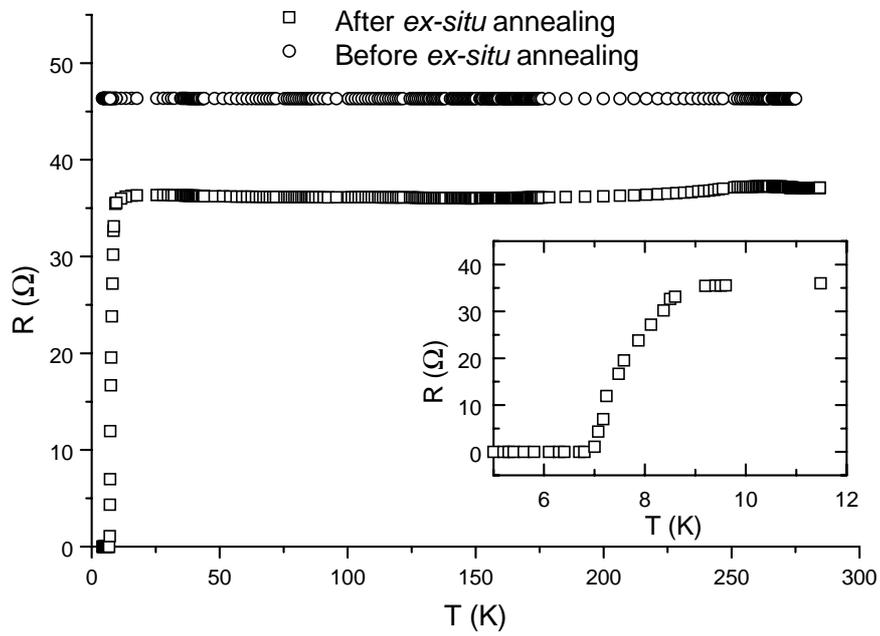

(b)

Figure 3

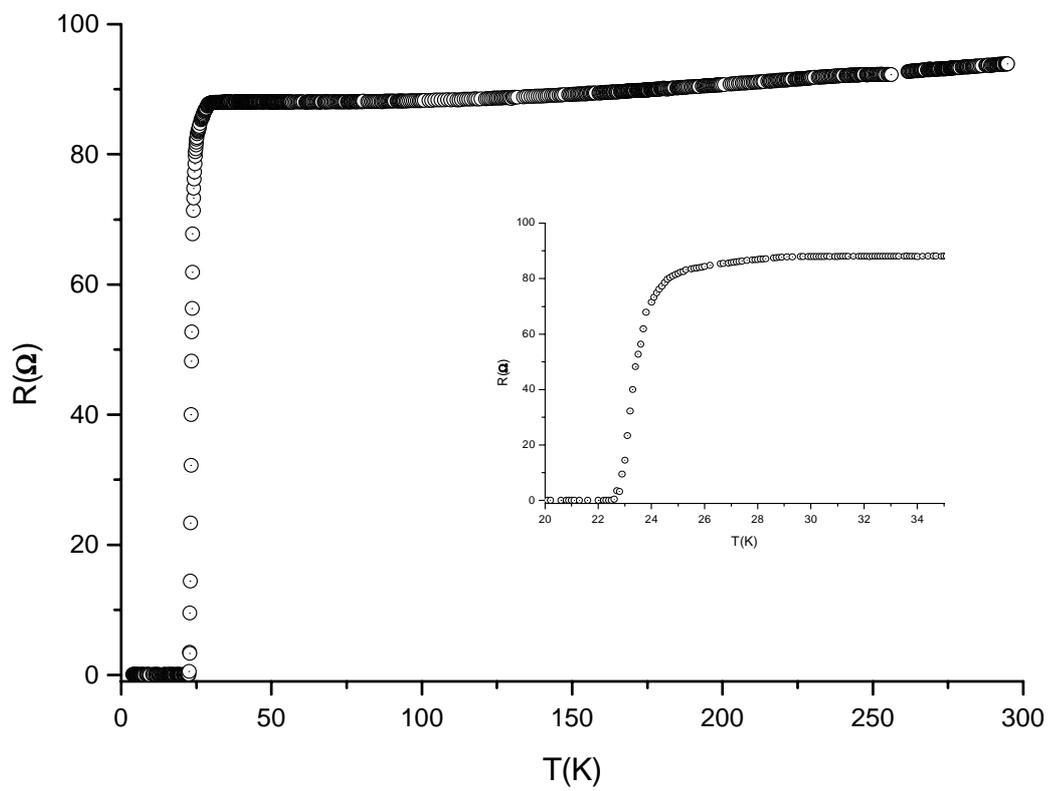

Figure 4